\documentclass[sigconf,screen] {acmart}
\usepackage[table]{xcolor}
\usepackage{graphicx}
\AtBeginDocument{%
  }

\setcopyright{acmlicensed}
\copyrightyear{2026}
\acmYear{2026}
\acmConference[CHI-Tools for Thought Workshop]{April,
  2026}{Barcelona, Spain}

\begin{document}

\title[Software Engineers' Cognitive Engagement]{``I'm Not Reading All of That'': Understanding Software Engineers' Level of Cognitive Engagement with Agentic Coding Assistants}


\author{Carlos Rafael Catalan}
\affiliation{%
  \institution{Samsung R\&D Institute Philippines}
  \city{Manila}
  \country{Philippines}}
\email{c.catalan@samsung.com}

\author{Lheane Marie Dizon}
\affiliation{%
  \institution{Samsung R\&D Institute Philippines}
  \city{Manila}
  \country{Philippines}}
\email{lm.dizon@samsung.com}

\author{Patricia Nicole Monderin}
\affiliation{%
  \institution{Samsung R\&D Institute Philippines}
  \city{Manila}
  \country{Philippines}}
\email{p.monderin@samsung.com}

\author{Emily Kuang}
\affiliation{%
  \institution{York University}
  \city{Toronto}
  \country{Canada}
}
\email{ekuang@yorku.ca}

\renewcommand{\shortauthors}{Catalan et al.}

\begin{abstract}
Over-reliance on AI systems can undermine users' critical thinking and promote complacency, a risk intensified by the emergence of agentic AI systems that operate with minimal human involvement. In software engineering, agentic coding assistants (ACAs) are rapidly becoming embedded in everyday development workflows. Since software engineers (SEs) create systems deployed across diverse and high-stakes real-world contexts, these assistants must function not merely as autonomous task performers but as Tools for Thought that actively support human reasoning and sensemaking. We conducted a formative study examining SEs’ cognitive engagement and sensemaking processes when working with an ACA. Our findings reveal that cognitive engagement consistently declines as tasks progress, and that current ACA designs provide limited affordances for reflection, verification, and meaning-making. Based on these findings, we identify concrete design opportunities leveraging richer interaction modalities and cognitive-forcing mechanisms to sustain engagement and promote deeper thinking in AI-assisted programming. 

\end{abstract}

\maketitle

\section{Introduction}

Recent advances in generative AI systems have brought about immense productivity gains in human-AI co-creative work. In software engineering in particular, agentic coding assistants (ACAs) such as Cline \cite{cline}, Claude Code \cite{claude-code}, and Codex \cite{codex} increasingly act as autonomous collaborators that generate, modify, and reason about code \cite{concode, test-assert-code-gen}. 
While these systems demonstrably accelerate development workflows, their growing agency raises critical questions about how they shape human cognition during complex problem-solving tasks.

Prior research has shown that over-reliance on AI systems can negatively affect core cognitive capabilities, such as decision-making \cite{sabharwal2023artificial, ahmad2023impact}, critical thinking \cite{duhaylungsod2023chatgpt}, and problem-solving \cite{duhaylungsod2023chatgpt}. These capabilities are especially crucial in software engineering, where developers must evaluate trade-offs, reason about correctness, and anticipate failure modes. 
At the same time, large language models (LLMs) that power these ACAs remain prone to hallucinations \cite{gao2022comparing} and algorithmic biases \cite{mbalaka2023epistemically}. 

Framed through the lens of \textbf{AI as ``Tools for thought,''} these tensions point to a design challenge rather than a purely technical one: how might ACAs be designed and used in ways that protect and augment human cognition, rather than displacing it? Understanding this requires moving beyond performance metrics to examine how developers cognitively engage with ACAs, how they make sense of AI-generated suggestions, and how they reflect on and evaluate these interactions during real work.

In this workshop contribution, we investigate software engineers’ cognitive engagement when working with agentic coding assistants to inform design and usage strategies for GenAI as tools for thought. Specifically, we ask:
\textit{\textbf{RQ1}: How cognitively engaged with the task are software engineers when working with agentic coding assistants?}, and 
\textit{\textbf{RQ2}: How did they recall, understand, analyze, and evaluate different aspects of their interaction with an ACA?}

We conducted a formative study with four software engineers whose professional experience ranged from less than 1 year to more than 10 years. 
Participants performed a single code generation task with an ACA, specifically Cline \cite{cline}. 
These tasks were designed and categorized using Bloom's Taxonomy as a theoretical framework for examining cognitive engagement \cite{bloom1956taxonomy}
Following the code generation task, participants self-reported their level of cognitive engagement through a survey assessing their reasoning, attention, and reflection during the task.
Our findings reveal two key patterns. First, participants' cognitive engagement declined as tasks progressed. Second, they were primarily focused on achieving correct outputs, paying little attention to the underlying process that produced them. 
Together, these behaviors led to shallow engagement, limiting participants’ ability to accurately recall, understand, analyze, and evaluate important details of the task.  
To address this decline in engagement, we propose design considerations for ACAs aimed at sustaining cognitive engagement and encouraging critical thinking beyond merely reaching correct solutions. 

\section{Background}
\subsection{Bloom's Taxonomy as Measurement of Cognitive Engagement}

Bloom’s Taxonomy conceptualizes cognition as a hierarchy of cognitive processes that vary in complexity and depth, progressing from basic forms of information processing to more advanced forms of reasoning and knowledge construction \cite{bloom1956taxonomy,anderson2001taxonomy}. At the lower end of this hierarchy, processes such as remembering and understanding involve the recognition and interpretation of information, whereas higher-level processes, such as analyzing and evaluating, require individuals to synthesize that information to reason about structures and relationships \cite{krathwohl2002revision}. 
In this taxonomy, higher-order cognition depends on the successful operation of foundational processes such that complex reasoning and judgment presuppose accurate recall and meaningful comprehension of relevant information, as well as coherent mental representations of task elements and their relationships \cite{forehand2010bloom}. 



This framework provides a conceptual basis for examining cognitive engagement in interactions with ACAs by characterizing engagement in terms of distinct cognitive processes. As ACAs increasingly assume planning and decision-making functions, task success alone offers limited insight into the underlying cognitive engagement of users. As such, cognitive engagement must instead be examined through the cognitive operations users employ while interacting with an ACA. Within this framework, levels such as remembering, understanding, analyzing, and evaluating correspond to qualitatively different forms of engagement, ranging from recall of interaction information to higher-order reasoning about the behavior and reliability of ACAs. This perspective allows us to assess both how cognitively engaged users are when interacting with ACAs and how that engagement is distributed across recall, comprehension, analytical reasoning, and evaluative judgment.

\subsection{Self-Report Surveys for Cognitive Engagement}

Self-report surveys serve as a reliable measurement tool for cognitive engagement \cite{greene2015measuring}, because it uses participants’ verbal responses to assess their current conscious cognitive state \cite{pekrun2020commentary}. These surveys have been used extensively in the education domain to measure students' motivations and learning goals \cite{entwistle1970relationships, entwistle2015understanding}. However, self-reporting is prone to memory biases. This is particularly true when the survey is administered at a later point in time and requires individuals' recall of key events from their autobiographical memory \cite{pekrun2020commentary}. For our use case, we use Bloom's Taxonomy as the framework for developing the questions for our survey. Specifically, we ask questions that would understand how well software engineers recalled, understood, analyzed, and evaluated various aspects of their interaction with Cline. To minimize memory biases, we administer this survey immediately after we ask them to perform a code generation task.

\subsection{Cognitive Load Theory}

Cognitive Load Theory suggests three types of cognitive demands on users during learning activities: intrinsic, extraneous, and germane load \cite{cognitive-load-theory}. Intrinsic load pertains to the amount of existing cognitive resources the user needs to use to understand complex information. Extraneous load refers to the cognitive resources the user needs to expend to comprehend irrelevant information resulting from poor material design. Germane load refers to the allocation of cognitive resources towards problem solving and metacognition \cite{krieglstein2022systematic, mutlu2019cognitive, sweller1998cognitive}. To enhance users' deep understanding of the learning material, it must be designed in a way that minimizes extraneous load, optimizes intrinsic load, and promotes germane load \cite{clark2023learning, mayer1998cognitive, mayer2003nine}.


\section{Method}

We conducted a user study to understand SEs' level of engagement with an ACA's output. We recruited four participants from a large software company in the Philippines. Each participant (P) represents a different category of professional SE experience: P1: less than 1 year, P2: 1-5 years, P3: 6-10 years, P4: more than 10 years. For the ACA, we selected Cline due to its agentic abilities of comprehensive planning and discussion with the SE, as well as invocation of external tools when executing the task \cite{Baumann_2025}. Cline is also the tool that participants were most familiar with. The agent is also open source \cite{cline}, affording us the flexibility to implement prototypes that encourage active user engagement for future work.


\subsection{Code Generation Task}

After going through the consent process, the moderator explained the study tasks. Each participant was presented with a laptop where the Visual Studio Code integrated development environment (IDE) is open, and the Cline agent is pre-loaded with the prompt for a Code Generation task:

\begin{quote}
{\ttfamily
Can you write a <programming language of your choice> script that checks all the Excel files in the folder and finds the one with a “dashboard” sheet? In the dashboard sheet, copy the values from column C to E. Then, generate another workbook that copies all the data from the current workbook, and names the new workbook’s sheet to whatever the name of the current workbook is.}
\end{quote}

This prompt was retrieved from DevGPT, a curated dataset of conversations between SEs and ChatGPT \cite{devgpt}.

Before beginning the code generation task, the moderator explained to the participants that they are afforded as much time as needed in interpreting the prompt and what they expect to be the correct output. They were also briefed on Cline's possible clarifying questions during the interaction, and that they should carefully analyze and use their best judgment when responding. Lastly, they were allowed to go back to any part of their conversation with Cline during the duration of the code generation task.

Once they understood the instructions, participants began the task by first indicating the programming language they were most comfortable working with before prompting Cline. Cline first began in `Plan' mode to provide them with an approach to solving the task. Once they thought the plan was sufficient, they then triggered `Act' mode to generate the code. During the `Plan' or `Act' modes, Cline may ask them some clarifying questions. 
The task ended once "Start New Task" appeared; they may do some reviewing before notifying the moderator that they were done. During the entire task, the moderator was present in the room to observe and record the participants' behavior. They were also allowed to think out loud, even though the moderator did not explicitly state the protocol.


\subsection{Survey Task}

Before proceeding with the survey, participants were instructed that they were not allowed to go back to the IDE. The survey was divided into five parts: (1) Years of Experience, (2) Recall, (3) Understanding, (4) Analyzing, and (5) Evaluation. The survey was deployed on Qualtrics.  Parts 2 to 5 contained questions aimed at measuring how well they \textbf{recalled}, \textbf{analyzed}, \textbf{understood}, and \textbf{evaluated} certain aspects of their interaction with Cline, following Bloom's Taxonomy. 

\subsection{Data Analysis}

We first cross-referenced each participant's responses to the recall questions with their corresponding working directory during the code generation task to determine if they correctly recalled various aspects of their interaction (e.g., the name of the folder they were working on). Then, two authors conducted a thematic analysis on the participants' short-answer responses and the moderator's observational notes \cite{step-by-step-ta}. From this analysis, we identified two key patterns, which we discuss in the next section.

\section{Findings}


We divide the interaction between the SE and Cline into three phases: planning, execution, and evaluation. The \textbf{planning} phase involves the SE understanding the prompt and the expected output, Cline creating a plan for the task, and the SE reviewing the plan. The \textbf{execution} phase involves Cline generating the actual code, running it, and evaluating it. The \textbf{evaluation} phase involves the SE evaluating both the generated source code and Cline's evaluation of that source code.

\subsection{SEs' Cognitive Engagement Declines as the Task Progresses}

In the context of our study, we observed a clear downward trend in the SEs' level of cognitive engagement as the task progressed, which provides some formative insights for RQ1.
Participants devoted the greatest cognitive effort during the planning phase, focusing on guiding the agent toward producing correct outputs. 
However, engagement dropped during the execution phase, where the volume of information presented by Cline often led to disengagement.
As a result, during the evaluation phase, SEs allocated minimal cognitive resources to verifying the output and largely neglected reviewing the underlying process that generated it.

\subsubsection{Software Engineers Allocated Most of Their Cognitive Resources Towards the Planning Phase}

We found that all participants allocated substantial cognitive resources in the Planning Phase, as evidenced by our observations that they were all actively engaged with the task at the beginning, especially when trying to comprehend the prompt. For example, P3 asked out loud: \textit{What is the context of this?}. P1, P2, and P4 took the time to verify the files in the working directory before prompting Cline.

Once Cline generated the plan, most participants went back through the conversation to read it. P1, P2, and P4 actively read the conversation by going back and forth between it and the generated plan, or slowly scrolling through it. This is in contrast to P3, who just skimmed the conversation to save time. 
We hypothesize that SEs attributed importance to tasks in relation to planning and orchestration of the agent to ensure that the agent understood the requirements to increase the probability of correctly performing the task. This may be a form of germane cognitive load, which led the participants to be meticulous in answering Cline's clarifying questions and reading through the conversation.

\subsubsection{Information Overload During Cline’s Execution Phase}

During Cline's execution phase, we observed that participants' engagement decreased while Cline was executing its process of code generation and tool invocation. Some of them (P2, P3) looked away from the laptop and towards the other parts of the room, and only when Cline provided results or clarification did the participants go back to the task. However, some participants did not necessarily engage with the results, as evidenced by P4's comment: \textit{"I'm not reading all of that"}. We also observed that other participants (P2, P3) were quick to prompt Cline for the next steps, even though there was still a considerable amount of information on the screen that had just been generated. 
Participants perceived this “information dump” as extraneous cognitive load, which overwhelmed their attention and reduced meaningful engagement with the task. Moreover, Cline’s reliance on text-only communication may have further amplified this extraneous load by requiring users to parse dense, unstructured information without additional visual or interactive support.

\subsubsection{Participants were More Concerned with Output Evaluation Rather than Process Evaluation} \label{evaluation-phase}


In the evaluation phase, our findings show that the four participants engaged more with the output than the process, as evidenced by the following responses: \textbf{(P1)}: \textit{"It generated my desired output"};  \textbf{(P2)}: \textit{"I trust Cline"}, \textbf{(P3)}: \textit{"It worked"}, and \textbf{(P4)}: \textit{"I think what cline wrote was well-documented and readable enough, so I felt that changes were not necessary."}. Further evidence for this finding was also observed with P4. During the task, he reflected on his expectations and evaluated that there were inconsistencies in the output, so he prompted Cline with clearer instructions until it gave him the correct results.

Once Cline successfully generated the correct Excel file, participants chose not to make further changes to the code. In this task, the Excel file served as the final output, while the code constituted the underlying process. Both required intrinsic cognitive effort to evaluate; however, verifying the output demanded substantially fewer cognitive resources than reviewing the code itself. When the output appeared correct, participants treated the task as complete (as observed in P2 and P3). Only when errors occurred did they engage in deeper process-level evaluation and revision (as observed in P4).
These behaviors suggest that software engineers adopt a \textit{greedy allocation strategy} for cognitive resources, prioritizing the lowest-effort checks that confirm task completion, and engaging in deeper reasoning only when surface-level validation fails.

\subsection{Participants Only Recalled, Understood, Analyzed, and Evaluated the "Happy Path"}

We offer some preliminary insights for RQ2. In the context of our study, participants appeared to recall, understand, analyze, and evaluate primarily the "happy path," that is, the sequence of steps leading to the correct output. This tendency to focus on the output rather than the process may have led participants to overlook potentially critical issues, which was reflected in the survey results.
In terms of recall, \textbf{none} of the participants were able to correctly answer how many functions the generated script had. For understanding, only \textbf{half} of them understood what the first function did and could reliably provide a concise summary for it. Likewise, only \textbf{half} were able to adequately analyze the code and felt confident that it could handle edge cases. 

These responses suggest that participants' evaluations may have been too narrow in scope, potentially overlooking unexpected or edge-case scenarios. Consider, for instance, a scenario in which an ACA is powered by an LLM susceptible to vulnerabilities such as poisoning and backdoor attacks \cite{trojan-puzzle, poisoning-vulnerabilities}. In such cases, SEs may need to recall, understand, analyze, and evaluate even seemingly minor aspects of their task to reduce the risk that malicious code goes undetected.

\section{Design Opportunities for ACAs}



\subsection{Sustain Engagement by Communicating Beyond Text}

During the ACA's execution phase, participant engagement appeared to be at its lowest. We hypothesize that the volume of text generated in real-time by the ACA may exceed what SEs can reasonably process. This limitation may be partly attributable to current ACAs relying solely on text as their communication medium. Faced with large amounts of streaming text, SEs may be inclined to skim rather than read carefully, potentially causing them to miss critical information relevant to their task.

To mitigate this, we propose to add visualization and alternative modalities, such as voice, to ACAs. Visualizations have long been an indispensable tool for human-centered design due to their ability to synthesize and communicate complex information \cite{ward2010interactive} for the users \cite{ware2019information, munzner2025visualization}. We envision an ACA that communicates its plans and reasoning in a presentable format, such as flowcharts, graphs, and mind maps. Previous work has also shown state-of-the-art voice AI systems that capture natural, human-like speech synthesis act as favorable social cues that elicit positive social responses from users \cite{beege2023emotional, liew2023alexa, liew2020does, schneider2022cognitive}. 
Studies by \citet{liew2025cognitive} and \citet{two-talking-heads} showed that the implementation of multiple AI voice technologies in learning systems resulted in significantly lower perceived cognitive load of the information presented, and improved retention and recall of the information by the user.

\subsection{Cognitive Forcing Designs to Encourage Critical Thinking}

In our study, participants' responses centered on the happy path, with few instances of extended recall, understanding, analysis, and evaluation of the ACA’s process and output. This suggests some over-reliance on the ACA for handling all possible edge cases. This observation aligns with the dual-process cognitive theory of \textit{System 1} thinking, where humans employ shortcuts (in this case, analyzing the happy path) in decision making to minimize use of cognitive resources \cite{kahneman2011thinking, wason1974dual}.

Participants' prior experience with Cline and understanding of Cline's capabilities may have encouraged them to stay on \textit{System 1} thinking. Designing systems that encourage users to switch to \textit{System 2} or analytical thinking is a long-standing challenge in human-AI collaboration research because it is more effort-intensive on the part of the user. Previous works on \textit{cognitive forcing designs} by \citet{to-trust-or-think} and \citet{ghosh2026experimental} show these to be a promising approach to encourage \textit{system 2} thinking.

Thus, we argue for such designs to be extended to ACAs. These designs are interventions that are applied during the AI's decision-making to disrupt its reasoning. This, in turn, \textit{forces} the user to perform analytical thinking in the task \cite{lambe2016dual}.  This "slowing down" of the AI's decision has been shown to greatly increase user accuracy of the assessment towards it \cite{slow-algorithm}. This finding aligns with another work done by \citet{kuang-ux-evaluation} in the domain of usability testing. In their work, AI's suggestions for usability improvements were only shown after the user was able to critically analyze the usability test video demonstration. This resulted in an improved perception of efficiency and trust towards the AI suggestions as it aligned with their desire to initially draw from their own expertise to address the task, and simply use the AI's analysis as a verification tool \cite{kuang-ux-evaluation}.

\section{Limitations, Future Work, and Conclusion}

Our work contains several limitations. We plan to continue this study by recruiting more participants, and add a software engineering task that is much more open-ended that would reward cognitive engagement. We also want to have an even deeper understanding of cognitive engagement to strengthen our design considerations. In addition to self-report surveys, we also plan to involve eye-tracking software to determine points of attention or lack thereof.

In conclusion, we present a formative study examining SEs' level of cognitive engagement when interacting with ACAs for code generation tasks. We found a concerning decline in cognitive engagement with the task as the interaction with the ACA progressed, which led our participants to overlook critical details.
To support the development of correct and safe real-world software, we propose design considerations for ACAs that sustain cognitive engagement and encourage critical thinking throughout the workflow, including multimodal interactions (e.g., visualizations and voice), and cognitive forcing designs. 
These design considerations are grounded in existing literature as well as our framing of ACAs as pair programmers. 
Drawing from established pair-programming practices, where collaboration is inherently multimodal and includes instructional scaffolding, we argue that ACAs should similarly support rich communication and intentionally provoke reflection rather than immediately providing solutions.

\bibliographystyle{ACM-Reference-Format}
\bibliography{sample-base}

\section{Appendices}

\subsection{Survey Questions}

\begin{enumerate}
  \item[\textbf{Q1}] (Single Choice) How many years of professional software engineering experience do you have?
  \begin{itemize}
     \item < 1 year
     \item 1-5 years
     \item 6-10 years
     \item > 10 years
   \end{itemize}
  \item[\textbf{Q2}] [Recall] What was the name of the folder/directory you were working on? (no need for the entire path, the folder name would be sufficient)
  \item[\textbf{Q3}] [Recall] What was the name of the first file created?
  \item[\textbf{Q4}] [Recall] How many functions/methods did the generated script have?
    \item[\textbf{Q5}] [Understand](Single Choice) "I can reliably provide a concise summary for the first function in the generated script"
    \begin{itemize}
     \item Strongly agree
     \item Somewhat agree
     \item Neither agree nor disagree
     \item Somewhat disagree
     \item Strongly disagree
   \end{itemize}
   \item[\textbf{Q6}] [Understand] (Single Choice) "I can reliably provide a concise summary for the last function in the generated script"
    \begin{itemize}
     \item Strongly agree
     \item Somewhat agree
     \item Neither agree nor disagree
     \item Somewhat disagree
     \item Strongly disagree
   \end{itemize}
   \item[\textbf{Q7}] [Analyze] (Single Choice) "I can reliably determine the order of functions/methods that will be called in the script"
    \begin{itemize}
     \item Strongly agree
     \item Somewhat agree
     \item Neither agree nor disagree
     \item Somewhat disagree
     \item Strongly disagree
   \end{itemize}
   \item[\textbf{Q8}] [Analyze] (Single Choice) The script is able to handle a case when there are no excel files in the working directory
    \begin{itemize}
     \item Yes
     \item No
     \item Unsure
   \end{itemize}
   \item[\textbf{Q9}] [Evaluate] (Single Choice) Did you make some of your own manual adjustments with the script?
    \begin{itemize}
     \item Yes
     \item No
   \end{itemize}
   \item[\textbf{Q10}] [Evaluate] (Open Ended) Why did you/did you not make any more adjustments?
\end{enumerate}

\subsection{Survey Results}

\begin{table}
\begin{tabular}{|| p{5em} | p{18em} ||} 
 \hline
 Participant & Response\\ 
 \hline\hline
 P1 &  It generated my desired output  \\ 
 \hline
 P2 &  I trust Cline  \\ 
 \hline
 P3 &  It worked  \\ 
 \hline
 P4 &  I think what Cline wrote was well-documented and readable enough, so I felt that changes were not necessary.  \\ 
 \hline
\end{tabular}
\caption{Participants' Responses to Q10: \textit{Why did you/did you not make any more adjustments (with the code)?}}
\label{table:1}
\end{table}

\begin{table}
\begin{tabular}{|| p{10em} | p{2.5em} | p{2.5em} | p{2.5em} | p{2.5em} ||} 
 \hline
 Question & P1 Correct? & P2 Correct? & P3 Correct? & P4 Correct?\\ 
 \hline\hline
 Q1 What was the name of the folder you were working on? & \cellcolor[HTML]{74f763} Yes & \cellcolor[HTML]{74f763} Yes & \cellcolor[HTML]{f24e4e} No & \cellcolor[HTML]{f24e4e} No  \\ 
 \hline
 Q2 What was the name of the first file created? & \cellcolor[HTML]{74f763} Yes & \cellcolor[HTML]{74f763} Yes & \cellcolor[HTML]{f24e4e} No & \cellcolor[HTML]{74f763} Yes  \\ 
 \hline
 Q3 How many functions/methods did the generated script have? & \cellcolor[HTML]{f24e4e} No & \cellcolor[HTML]{f24e4e} No & \cellcolor[HTML]{f24e4e} No & \cellcolor[HTML]{f24e4e} No  \\ 
 \hline
\end{tabular}
\caption{Participants' Responses to the Recall Questions. Their responses were cross-referenced with the directory they worked on to verify if their responses were correct.}
\label{table:2}
\end{table}


\includegraphics[scale=0.4]{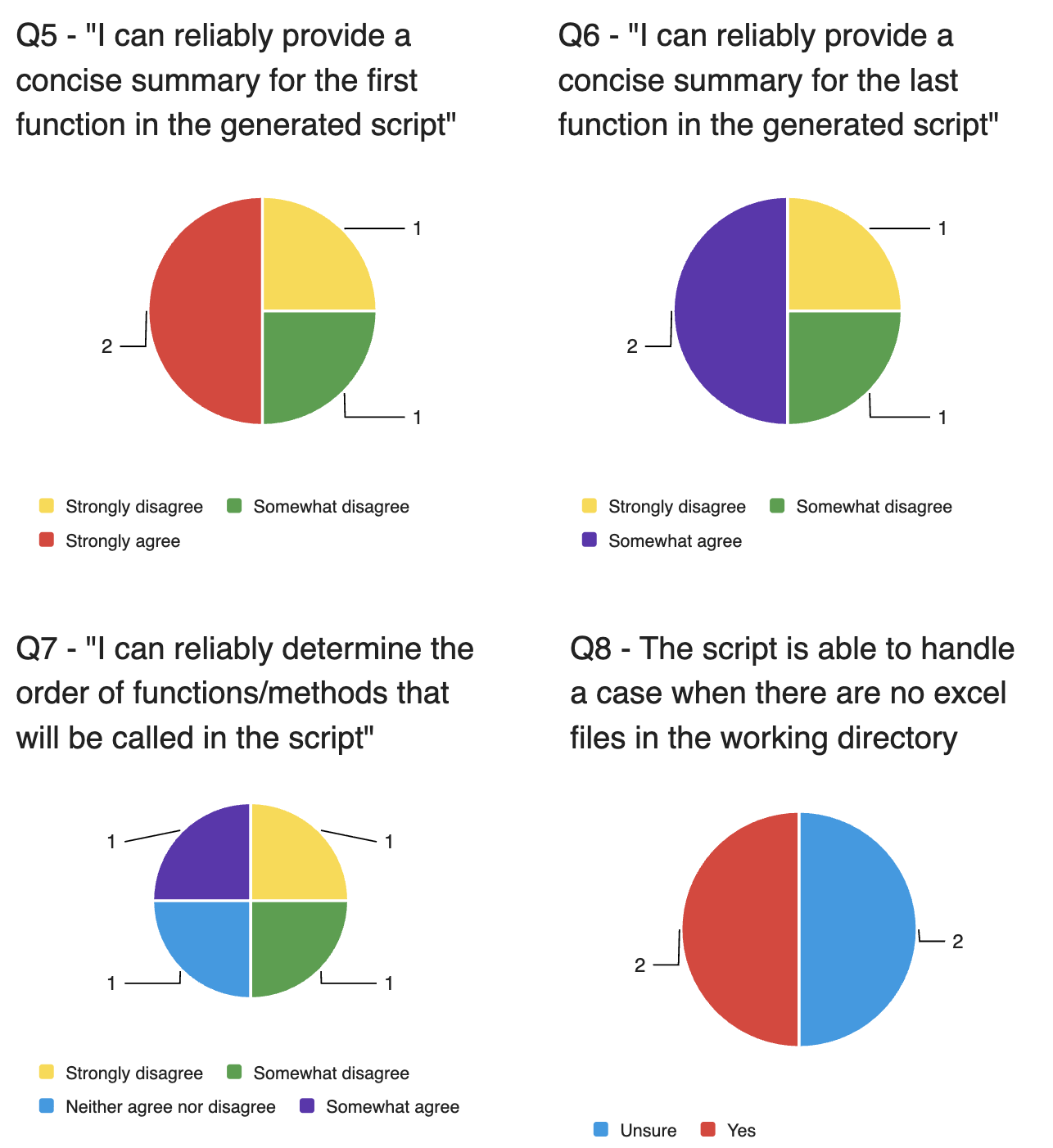}
\includegraphics[scale=0.4]{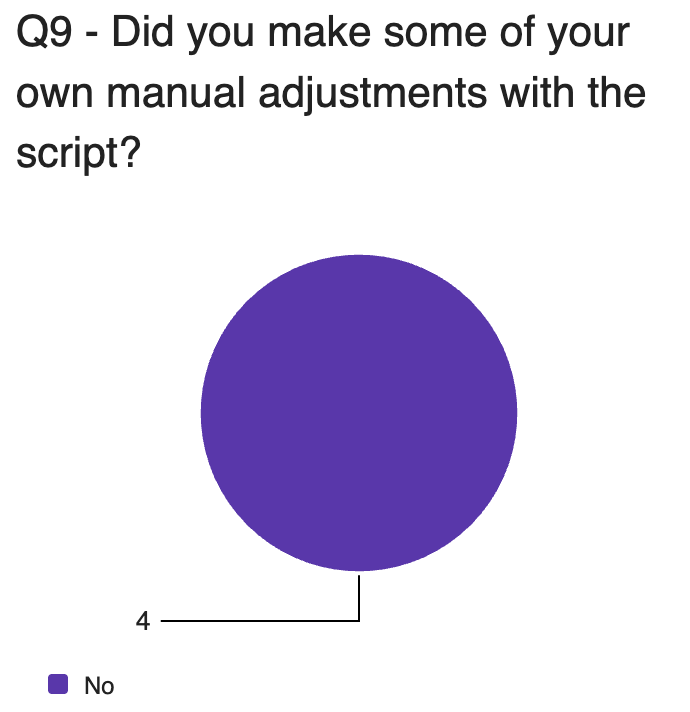}

\end{document}